\documentclass[12pt,a4paper]{article}
\usepackage{amsmath,amssymb,graphicx}

\begin{document}

\textwidth=135mm
 \textheight=200mm
\begin{center}
{\bfseries Twistor structures and boost-invariant solutions to field equations}
\vskip 5mm
V. V. Kassandrov$^{\dag}$~\footnote{E-mail: vkassan@sci.pfu.edu.ru}, J. A. Rizcallah$^\ddag$~\footnote{E-mail: joeriz68@gmail.com} and N. V. Markova$^\S$~\footnote{E-mail: n.markova@mail.ru}
\vskip 5mm
{\small {\it $^\dag$ Institute of Gravitation and Cosmology, Peoples' Friendship University of Russia, 117198, Moscow, Russia}} \\
{\small {\it $^\ddag$ Lebanese University, School of Education, Beirut, Lebanon}}\\
{\small {\it $^\S$ Department of Applied Mathematics, Peoples' Friendship University of Russia, Moscow, Russia}}\\






	
\end{center}
\vskip 5mm
\centerline{\bf Abstract}
\noindent
{\small We give a brief overview of a non-Lagrangian approach to field theory based on a generalization of the Kerr-Penrose theorem and algebraic twistor equations. Explicit algorithms for obtaining  the set of fundamental (Maxwell, $SL(2,\mathbb C)$-Yang-Mills, spinor Weyl and curvature) fields associated with every solution of the basic system of algebraic equations are reviewed. The notion of a boost-invariant solution is introduced, and the unique axially-symmetric and boost-invariant solution which can be generated by twistor functions is obtained, together with the associated fields. It is found that this solution possesses a wide variety of point-, string- and membrane-like singularities exhibiting nontrivial dynamics and transmutations.}




\section{Introduction}
\label{intro}
Twistors $\bf W$ of the Minkowski space-time $\bf M$ can be defined as the pairs of 2-spinors $\xi,\tau$ linked by the Penrose {\it incidence relation} $\tau = Z\xi$ with the points of $\bf M$ represented by an Hermitian matrix of coordinates $Z=Z^+$~\cite{pr}. Since the incidence condition remains invariant under the shift $Z\mapsto Z+ c \xi^+ \xi$, $c \in \mathbb R$, any point in twistor space actually corresponds to a {\it light-like ray} in $\bf M$. Therefore, twistors are closely related to null geodesic congruences on $\bf M$ and reveal the ``hidden''  twistor structure of relativistic field equations. 

Actually, a wide class of solutions of fundamental field equations possesses a null {\it shear-free} congruence (NSFC) which to a great extent determines their structure. Particularily, the Kerr and Kerr-Newman solutions in GTR can be represented in the {\it Kerr-Shild form} \begin{equation}\label{KS}
 g_{\mu\nu} =\eta_{\mu\nu} + h k_\mu k_\nu,
 \end{equation}
 where $k_\mu$ is a null 4-vector field whose lines are tangent to a corresponding NSFC.   

On the other hand, the structure of NSFC on $\bf M$ (or its Riemannian deformation (\ref{KS})) was completely described through the celebrated {\it Kerr-Penrose theorem}~\cite{dks,pr}. Specifically, any (projective) 2-spinor field $\xi$ which is a continious solution of the equation 
\begin{equation}\label{KerrN}
\Pi({\bf W}) = \Pi(\xi, Z\xi) =0,
\end{equation}
where $\Pi$ is an arbitrary {\it homogeneous} function of (three projective) twistor variables, defines the null 4-vector field $k_\mu = \xi_A \xi_{A^\prime}$ which necessarily is a tangent field of a NSFC. Moreover, all the NSFC on $\bf M$ can be generated in such a purely algebraic way. 

 Remarkably, an invariant form of the twistor equation (\ref{KerrN}), 
 \begin{equation}\label{GenSol}
 \Pi^{(C)} (\xi, Z\xi) = 0,~~C=1,2
 \end{equation}      
(now $\Pi^{(C)}$ are two {\it arbitrary} functions of four twistor variables $\xi,\tau=Z\xi$)
arises, in the framework of so-called {\it algebrodynamical (AD) approach}~\cite{kass92,kass95,kassriz16}.  In fact, (\ref{GenSol}) represents the general solution of the conditions of {\it biquaternionic $\mathbb B$-differentiability}~\cite{kass92,kass95} which are, in a sense, a natural generalization of the Cauchy-Riemann conditions in complex analysis. The form (\ref{GenSol}) reveals many ``hidden'' properties of the NSFC and, in particular, allows for a direct definition of a number of fundamental (both gauge and spinor) fields: complex Maxwell field, $SL(2,\mathbb C)$ Yang-Mills field, Weyl 2-spinor field and others~\cite{kassriz17}. For all of them, the corresponding vacuum equations (with rich structure of field singularities) are necessarily fulfilled as a direct consequence of the sole condition (\ref{GenSol}). 

Thus, in the AD approach one regards the equations of $\mathbb B$-differentiability as the primary field equations while Maxwell, Yang-Mills and Einstein equations -- as mere definitions of the singular sources-particles of the EM and gravitational fields respectively. Close relationship with twistor structures allows then one to solve the basic AD equations and the associated gauge and spinor equations purely algebraically. Remarkably, all the generated solutions to the free Maxwell equations have an integer electric charge for each of the isolated singularities~\cite{Sing}.

\section{Differential equations, gauge symmetry and fields defined by twistor constraints}
\label{over}

By direct differentiation of the algebraic constraint (\ref{GenSol}) one obtains the following equations~\footnote{We use standard 2-spinor notation, whereby upper case Latin indices range and sum over $0$ and $1$, and are raised and lowered by the symplectic spinors  $\epsilon^{AB}$ and $\epsilon_{AB}$. Notation $\nabla_{AA^\prime}$ means ordinary partial differentiation w.r.t corresponding coordinate $Z^{AA^\prime}$.} for the principal spinor $\xi_A,~(A=0,1)$: 
\begin{equation}\label{SF}
\xi^{A^\prime} \nabla_{A A^\prime} \xi_{B^\prime} = 0,
\end{equation} 
which are just an invariant form of the defining conditions of NSFC (following from (\ref{KerrN}), respectively),  
\begin{equation}\label{SFC}
\xi^{A^\prime} \xi^{B^\prime} \nabla_{A A^\prime} \xi_{B^\prime} = 0. 
\end{equation}
The rigid form (\ref{SF}) is, in the turn, equivalent to the following system of equations:
\begin{equation}\label{GSE}
\nabla_{A A^\prime} \xi_{B^\prime} = \Phi_{B^\prime A} \xi_{A^\prime}
\end{equation}
for some 2-spinor $\xi_{A^\prime}(Z)$ and 4-vector $\Phi_{B^\prime A}(Z)$ fields.

Equations (\ref{GSE}) possess "weak" gauge symmetry of the form
\begin{equation}
\label{gauge}
\xi_{A^\prime} \to \alpha(Z)\xi_{A^\prime},\ \
\Phi_{B^\prime A} \to \Phi_{B^\prime A} + \nabla_{A B^\prime} \ln \alpha(Z),
\end{equation}
with the function $\alpha(Z)$ implicitly depending on $Z$ through the components of the twistor  $\{\xi_{A^\prime}, \tau^{A} := Z^{AA^\prime}\xi_{A^\prime}\}$, i.e. $\alpha(Z) =\alpha (\xi_{A^\prime}, \tau^{A})$. 

The consistency of (\ref{GSE}) implies {\em self-duality} of the ``curvature" 2-form associated with the matrix 1-form 
\begin{equation}\label{omega}
\Omega_{B^\prime}^{~~~A^\prime} = \Phi_{B^\prime A}dZ^{A A^\prime}.
\end{equation}
  This, in turn, leads to Maxwell and the $SL(2, \mathbb C)$ Yang-Mills vacuum equations for the trace and trace-free parts of the matrix $\Omega$. As shown in~\cite{kassriz17}, a Weyl 2-spinor field $\phi_{A}(Z)$ can also be constructed out of the components of $\Phi_{B^\prime A}(Z)$. Moreover, all the named fields possess a common singularity, corresponding to spacetime events where system (\ref{GenSol}) admits multiple solutions, given by 
\begin{equation}\label{sing}
\det \Vert \frac{d \Pi^{(C)}}{d\xi_{A^\prime}} \Vert = 0.
\end{equation}
	
\section{The fundamental solution}\label{fix}
Using the symmetry (\ref{gauge}), we may set $\xi_{0^\prime}=1$. In this gauge,  systems (\ref{SF}) and (\ref{SFC}) take the same form 
\begin{equation}\label{Geq}
\nabla_\omega G=G\nabla_{u}G,~~~\nabla_{v}G=G\nabla_{\bar \omega}G,
\end{equation}
for the gauge invariant $G:=\xi_{1^\prime}/\xi_{0^\prime}$, where use has been made of the canonical variables
\begin{equation}\label{zcoor}
Z^{AA^\prime}=
  \begin{pmatrix}
    u & \omega \\
    \bar \omega & v
  \end{pmatrix}\equiv
  \begin{pmatrix}
    t+z  & x-iy \\
    x+iy & t-z
  \end{pmatrix},
\end{equation}  
with $t$ and $\{x, y, z\}$ being the ordinary real time and Cartesian space coordinates. As a result, the constraints (\ref{KerrN}) or (\ref{GenSol})  reduce to 
\begin{equation}\label{Gsol}
\Pi(G, \tau^0, \tau^1)= \Pi(G,u+\omega G, \bar \omega+vG)=0,
\end{equation}
and the singular set condition (\ref{sing}) simplifies to
\begin{equation}\label{Gsing}
\frac{d\Pi}{dG}=0,
\end{equation}
with $\Pi$ being an arbitrary analytic function of three complex variables. It follows from (\ref{Geq}) that, for every solution of (\ref{Geq}), any $C^2$-function of $G$ satisfies both the eikonal and the wave equations~\cite{kassriz17}.   

As a consequence, Maxwell vacuum equations hold for the trace 
\begin{equation}\label{EMA}
A:=\frac{1}{2G}(\nabla_v G dv +\nabla_{\omega}G d\omega) 
 \end{equation}
of the matrix 1-form (\ref{omega}), which in the present gauge reduces to
\begin{equation}\label{Omega}
\Omega=
  \begin{pmatrix}
    0 & 0 \\
    dG -2GA & 2A
  \end{pmatrix}.
 \end{equation}
Moreover, the trace-free part of (\ref{Omega})  
$$
{\bf \Omega}=
  \begin{pmatrix}
    -A & 0 \\
    dG -2GA & A
  \end{pmatrix}
$$
 has a strength, ${\bf F} :=d {\bf \Omega} - {\bf \Omega} {\bf \Omega}$, of the form~\cite{diss}
\begin{equation}\label{vecF}
{\bf F} =
  \begin{pmatrix}
    -1 & 0 \\
    -2G & 1
  \end{pmatrix}F,
 \end{equation}
satisfying the Yang-Mills equation, with $F:=dA$ being the electromagnetic field strength.  It is worth noting that while the Yang-Mills field ${\bf F}$ is non-Abelian, by of the non-vanishing term ${\bf \Omega}{\bf \Omega}$, the Bianchi identity as well as the Yang-Mills equation for $\bf F$ are equivalent to their Abelian counterparts for $F$. Finally, the associated Weyl field $\phi_A$ is calculated using~\cite{kassriz17}
\begin{equation}\label{WeylG}
\phi_0:=\Phi_{1^\prime 0} =\frac{\nabla_{\omega} G}{G},~~\phi_1:=\Phi_{1^\prime 1} =\frac{\nabla_{v} G}{G}. 
\end{equation}
and on the solutions of (\ref{Geq}) satisfies the Weyl equations $\nabla_{\bar \omega} \phi_0 =\nabla_{u} \phi_1, ~ \nabla_v \phi_0 =\nabla_{\omega} \phi_1$.

By way of a simple, yet fundamental, example, let us consider the static solution of (\ref{Geq}), generated by the function $\Pi = G \tau^0 -\tau^1 -2bG$, whose explicit form is
\begin{equation}\label{fundG}
G = \frac{\bar \omega}{(z-b) \pm \sqrt{\rho^2+(z-b)^2}},
\end{equation}
where $\rho:=\sqrt{\omega\bar\omega} = \sqrt{x^2 + y^2}$ and $b = Const. \in \mathbb{C}$. The (real part of) self-dual electromagnetic field associated with (\ref{fundG}) is just the Coulomb field
\begin{equation}\label{fundE}
{\bf E} = \frac{1}{4(\rho^2+(z-b)^2)^{3/2}}\{x,y,z-b\}.
\end{equation}
Using (\ref{fundG}) and (\ref{fundE}), the corresponding Yang-Mills (\ref{vecF}) and Weyl (\ref{WeylG}) fields can be found, with the latter being
\begin{equation}\label{fundW}
\phi_0 = \frac{\mp G}{2\sqrt{\rho^2+(z-b)^2}},~~\phi_1= \pm \frac{1}{2\sqrt{\rho^2+(z-b)^2}}.
\end{equation}
 
For real $b$, all the physical fields are singular at the point charge located on the $z$-axis at $z=b$. In addition, the Yang-Mills and the Weyl fields admit a string-like singularity correponding to those of $G$,  
\begin{equation}\label{singfG}
\{\rho=0\} \cap \{z-b < 0 \}~ or~\{z-b > 0 \}, 
\end{equation}
for the first and second mode, respectively. This is a string that stretches along the $z$-axis from the very location of the point charge all the way to $z=- \infty$ or $z=+ \infty$. 

For complex $b$, on the other hand, all the associated physical fields are singular on a charged ring $\{z=\Re{b}\} \cap \{\rho=|\Im{b}|\}$, regardless of the choice of the branch of $G$ which now becomes two-valued. With the appropriate choice of the branch of $G$ it is possible to ensure that the ring is the only singularity of any one of the fields. The Maxwell and curvature fields (the latter defined by the metric (\ref{KS})) are now defined by the Kerr NSFC and correspond to the Kerr-Newman solution of the Einstein-Maxwell equations (yet, with a fixed ``elementary'' value of the singular Kerr ring)~\cite{kass92,Sing}.  

\section{Doubly invariant solutions and field singularities}
\label{sings}
Among all the solutions to (\ref{Geq}), of particular physical interest are those which give rise to fields invariant with respect to a geometric or internal group of symmetry. 
The class of solutions with {\it axially symmetric} electromagnetic and other fields is of special interest, as they may be associated with the structure of charged particles with spin. These fields may be defined as those for which the singular set as well as  scalar combinations of their components do not depend on the angle of rotation $\varphi$, say, in the $xy$ plane. 

Hence, to obtain solutions with axially symmetric fields from (\ref{Gsol}), one has to factor out the $\varphi$-dependence in $G$, so that $G(u,v,\omega, \bar \omega)=(\bar \omega/\rho) H(u,v,\rho) \equiv e^{\imath \varphi} H$. It follows then that $\tau^0 = u + \rho H$ and $\tau^1/G = (\rho+vH)/H$ are axial invariants, so that the constraint (\ref{Gsol}) reduces to the equation
\begin{equation}\label{axialsym}
\Pi(\tau^0,\tau^1/G)=\Pi(u + \rho H, \rho+vH)/H) =0 
\end{equation}
for an unknown axial symmetric function $H(t,z,\rho)$. 

Axial symmetry should have an obvious analog in relativity, namely {\it boost invariance}~\cite{kasstret}. While the former involves ordinary rotations in space, the latter entails hyperbolic rotations in spacetime. For the latter to be consistent with axial symmetry, it seems natural to consider boosts along the $z$-axis:
$$
t \to t \cosh{\theta} + z\sinh{\theta},\\
z \to t \sinh{\theta} + z\cosh{\theta},
$$
where $\theta$ is the hyperbolic angle. 

By analogy with the above, we can ensure boost-invariance of the associated fields by requiring the form-invariance of (\ref{Gsol}) and (\ref{Gsing}) with respect to hyperbolic rotations. The function $G$ should then be represented in the form  $G(u,v,\omega, \bar \omega)= (u/\sigma)K(\omega,\bar \omega,\sigma)$, where $\sigma:  =\sqrt{uv}=\sqrt{t^2-z^2}$, so as to make explicit the boost invariance of $\tau^0/G = (\sigma + \omega K)/K$ and $\tau^1 = \bar \omega + \sigma K$, and thus conclude for the reduced form of the function
\begin{equation}\label{boostinv}
\Pi(\tau^0/G,\tau^1)=\Pi(\sigma + \omega K)/K, \bar \omega + \sigma K)=0.
\end{equation}
 
Combining (\ref{axialsym}) and (\ref{boostinv}), we see that the function $\Pi$ must depend on the twistor variables through the unique combination $\tau^0 \tau^1/G$. Therefore, axially symmetric and boost-invariant electromagnetic fields are generated by the one and only functional form~\cite{kasstret}
\begin{equation}\label{joe}
\tau^0\tau^1 + b^2 G= 0,
\end{equation}
where $b$ is some complex constant.
Solving for $G$ in (\ref{joe}), we find (for one of the two modes):
\begin{equation}\label{G}
G = \frac{-(\rho^2 +\sigma^2+b^2)+\sqrt{\Delta}}{2\omega v},
\end{equation}
where $\Delta:=(\sigma^2-\rho^2+b^2)^2+4b^2\rho^2$.

For $b^2>0$, the function $G$ is singular at 
\begin{equation}\label{singG}
\{\rho=0\} \cap \{z^2 - t^2 \geq b^2\}.
\end{equation}
This represents a pair of strings that stretch along the $z$-axis from $z=\pm\sqrt{t^2+b^2}$ all the way to $z=\pm \infty$. The parameter $b$ determines the closest approach $2|b|$ between the strings' tips, which execute hyperbolic motion along the $z$-axis, arriving from $z=\pm \infty$ at past infinity then receding to $z=\pm \infty$ at future infinity.  

The  (real part of) self-dual electromagnetic field associated with (\ref{G}) is given by the following nonzero components~\cite{diss}:
\begin{equation}\label{emborn}
E_\rho = \frac{8b^2 \rho z}{\Delta^{3/2}},~ E_z =\frac{4b^2 M}{\Delta^{3/2}},~H_\varphi = \frac{8b^2 \rho t}{\Delta^{3/2}}, 
\end{equation}
where $M:=\sigma^2 +\rho^2 +b^2$.

For $b^2>0$, the electromagnetic field represents Born's well-known solution in electrodynamics~\cite{ror}, with point-like singularities (corresponding to $\Delta=0$) at the tips of the strings (\ref{singG}), i.e. $\{\rho=0\} \cap \{z=\pm \sqrt{t^2+b^2}\}$. It can be easily shown that the singularities of the electromagnetic field carry opposite ``elementary'' electric charges.

Moreover, being essentially the product of $G$ and the electromagnetic 2-form (see (\ref{vecF})), the Yang-Mills field ${\bf F}$, for the present case of $b^2>0$, is singular along the strings (\ref{singG}), including their tips. So is the Weyl 2-spinor, whose components are
\begin{equation} \label{weyl}
\phi_0 = \frac{\sigma^2-\rho^2+b^2-\sqrt{\Delta}}{2\omega \sqrt{\Delta}},\ \
\phi_1= \frac{\rho^2-\sigma^2+b^2-\sqrt{\Delta}}{2v\sqrt{\Delta}}.
\end{equation}
Thus, for $b^2>0$, all the fields share a pair of oppositely charged point-like singularities which execute hyperbolic motion along the $z$-axis. However, the fields ${\bf F}$ and $\phi_A$ admit additional string-like singularities that start at these electric charges and stretch to $z=\pm \infty$.           

Consider now the case $b^2=-a^2<0$. The common singular set for all physical fields ($\Delta=0$) is then given by 
\begin{equation}\label{tor}
z^2+(\rho-|a|)^2 = t^2.
\end{equation}
At $t=0$ the singularity starts out as a ring of radius $|a|$ in the $xy$-plane, inflates at the speed of light into a torus which self-intersects at the origin at $t=|a|$ and eventually turns into a sphere of very large radius ($\gg|a|$). By time reversal symmetry, for $t<0$ the singularity back-traces its evolution during the $t>0$ phase. Unlike the case $b^2>0$, the singular set (\ref{tor}) stays electrically neutral at all times. Incidentally, while the ``hole'' of the singular torus is open ($t<|a|$), the electromagnetic field inside the torus is ``dual'' to that in its exterior, where the electric field lines are topologically linked with the torus.  

Here too, it turns out that the Yang-Mills fields and the Weyl field (\ref{weyl}) have in common with $G$ an additional singularity, which is not part of the singular set (\ref{tor}) of the electromagnetic field. This singularity comes in two types: a pair of strings along the $z$-axis
\begin{equation}\label{str}
\{\rho=0\} \cap \{z^2 - t^2 \geq -a^2\},
\end{equation}
and a circular membrane parallel to the $xy$-plane
\begin{equation}\label{mem}
\{\rho \leq|a|\} \cap \{z = t\}.
\end{equation}         
The membrane (\ref{mem}) moves at the speed of light, whereas the tips of the strings (\ref{str}) (when they exist) have superluminal speeds. 

\begin{figure}[ht]
\vskip4cm
\includegraphics[angle=0,scale=0.32]{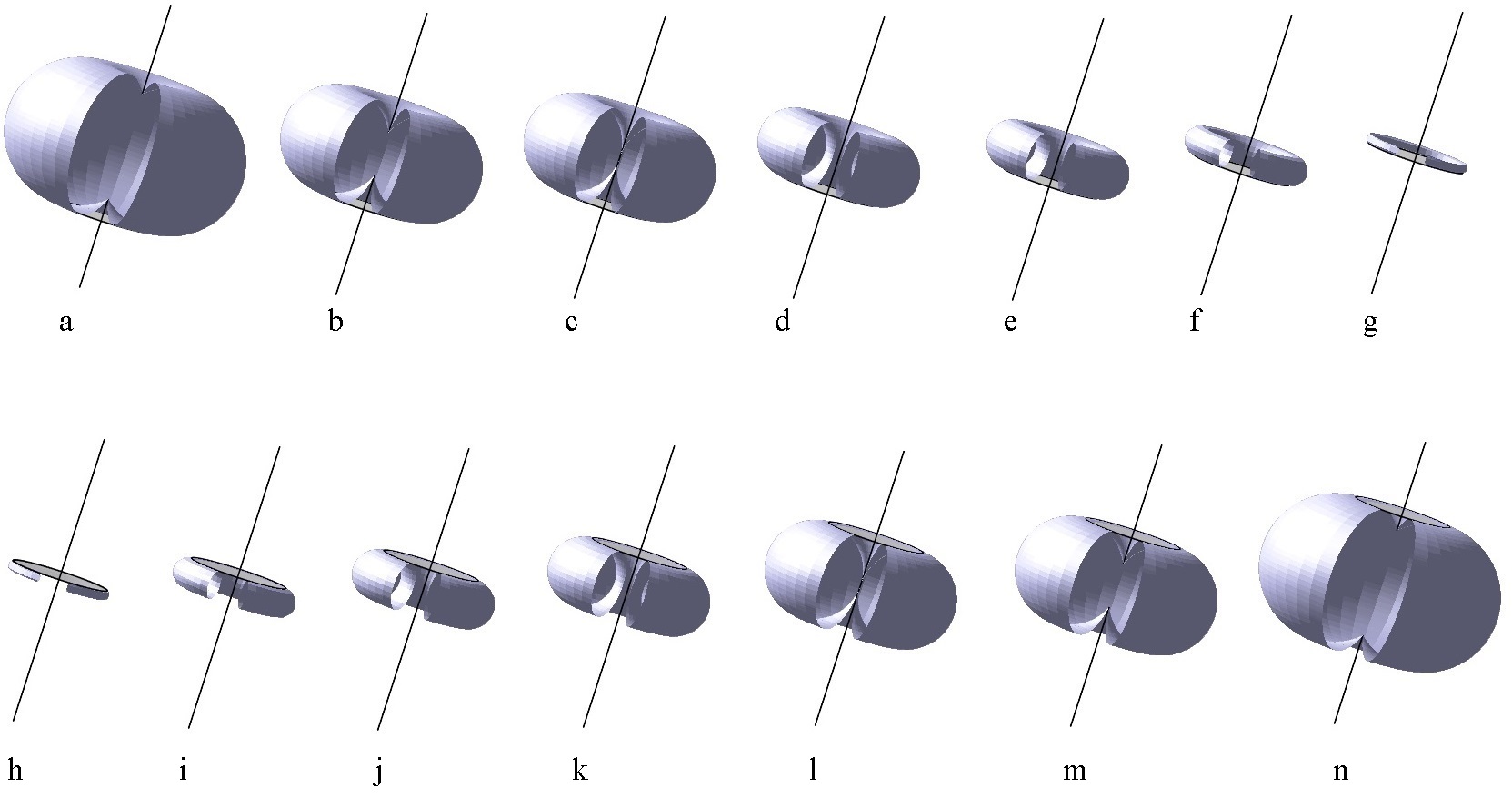}
\caption{{\small A depiction of the time evolution of the fields' singularities for imaginary values of the parameter $b$; the direction of time is from the upper left corner to the lower right one. The sliver disk and the dark straight line represent respectively the membrane-like (\ref{mem}) and string-like (\ref{str}) singularities of the Yang-Mills and Weyl fields. To enhance the figure, a little ``sector'' has been removed from the singular torus (\ref{tor}).}}
\label{pic1}
\end{figure}

For $|t|<|a|$, the  strings (\ref{str}) are merged together through the origin and stretch over the entire $z$-axis, while the membrane (\ref{mem}), being always in contact with the torus (\ref{tor}) around the ring $\{\rho=|a|\}\cap\{z=t\}$, moves along the $z$-axis (see Fig.~\ref{pic1} (d)-(k)). 

In particular, at $t=0$ the membrane crosses the $xy$-plane filling the interior of the common singular ring into which the singular torus degenerates then. At later times, $t>|a|$, the string rips at the origin and the resulting strands begin to retract along the negative and positive $z$-axes with their tips coinciding with the self-intersection points of the torus (\ref{tor}). Meanwhile, the membrane (\ref{mem}) continues to move at the speed of light ``riding'' on top of the torus around the ring $\{\rho=|a|\}\cap\{z=t\}$ (see Fig.~\ref{pic1} (l)-(n)). 
For $t<-|a|$, the evolution takes place in reverse: the strings merge at the origin as the membrane moves toward the origin ``sealing'' the bottom of the singular torus around the ring $\{\rho=|a|\}\cap\{z=t\}$ (see Fig.~\ref{pic1} (a)-(c)).                 

Finally, for the generic case of a complex parameter $b=b_0+ib_1, b_0, b_1 \in \mathbb{R}$, the function $G$ is double-valued and admits branch points at the common singular set $\Delta = 0$, which, in the present case, consists of a pair of oppositely charged Kerr-like rings, with equal radii, executing hyperbolic motion in opposite directions along the $z$-axis 
\begin{equation}\label{rin}
\{\rho =|b_1|\} \cap \{z=\pm\sqrt{t^2+b_0^2}\}.
\end{equation} 
On one of its sheets the function $G$ is nowhere singular, so neither the Yang-Mills fields nor the Weyl field associated with that branch have any additional singularities. 

\section{Conclusion}
\label{con}
The twistor framework provides an elaborate scheme of generating solutions to fundamental field equations. In the paper, we obtain the unique (up to an arbitrary complex parameter) axially-symmetric and boost-invariant solution to the basic system of twistor algebraic equations. We find that the physical fields generated by this solution exhibit a wide variety of bounded singularities, such as points or membranes, as well as unbounded string-like ones. These singularities may be electrically neutral or charged. In the latter case the value of charge is necessarily ``self-quantized''~\cite{Sing} being an integral multiple of that of the fundamental Kerr-like solution (\ref{fundE}). Such ``particle-like'' singularities seem to display a rich point-, string- and surface-like dynamics. In fact, for more complicated generating twistor functions the structure and dynamics of the singularity induced by solutions of (\ref {GSE}) can be far richer than presented here~\cite{Sing}.

\end{document}